\documentclass[aps,pra,10pt,superscriptaddress,twocolumn]{revtex4-2}
\usepackage{graphicx}
\usepackage{amsmath,amsthm,amssymb,dsfont}
\usepackage{mdframed}

\usepackage{orcidlink}
\newcommand{\ket}[1]{\left| #1 \right\rangle}
\newcommand{\bra}[1]{\left\langle #1 \right|}

\newcommand{\ketbra}[2]{\left|#1\right\rangle\hskip-1mm\left\langle#2\right|}

\usepackage{hyperref}

\begin{document}

\title{An operator-based bound on information and disturbance in quantum measurements}

\author{Hollis Williams}
\email{holliswilliams@hotmail.co.uk}
\thanks{Current address: Department of Mathematics and Statistics, University of Exeter, Exeter EX4 4QF, UK}
\affiliation{Graduate School of Advanced Science and Engineering, Hiroshima University, Kagamiyama 1-3-1, Higashi Hiroshima 739-8530, Japan}

\author{Holger F. Hofmann}
\email{hofmann@hiroshima-u.ac.jp}
\affiliation{Graduate School of Advanced Science and Engineering, Hiroshima University, Kagamiyama 1-3-1, Higashi Hiroshima 739-8530, Japan}

\begin{abstract} 
Quantum measurements can be described by operators that assign conditional probabilities to different outcomes while also describing unavoidable physical changes to the system. Here, we point out that operators describing information gain at minimal disturbance can be expanded into a set of unitary operators representing experimentally distinguishable patterns of disturbance. The observable statistics of disturbance defines a tight upper bound on the information gain of the measurement.

\end{abstract}

\maketitle

\section{Introduction}

It is well known that there is an information-disturbance tradeoff in quantum mechanics. Any information gain due to a measurement of the system also implies a disturbance of that system \cite{pekka, pekka2}.  In terms of operators, a quantum measurement necessarily changes states that are not eigenstates of the measurement operator.  However, the precise evaluation of this tradeoff using concepts of information theory is surprisingly difficult \cite{sch, sacchi, bus, horo, hall, busch}.  Existing literature employs tools from quantum information theory such as mutual information, quantum discord, and mean estimation fidelity to bound the tradeoff between information gain and the disturbance introduced to the system during measurement \cite{mutual, ren, sabe}. These tools tend to focus on the information content of quantum states, making it somewhat difficult to identify the role of the measurement process itself in the definition of the tradeoff. Here, we propose to shift the focus to the operators that describe the measurement process, introducing an analysis of the operator algebra that defines the information-disturbance tradeoff.
{\color{black} This means that the information gain is represented by the complete set of conditional probabilities that characterizes the measurement. We can then obtain a bound that is tighter than any bound based on a single information measure, demonstrating that the precise structure of measurement errors enters into the disturbance caused by the measurement interaction. Our result thus highlights the tight relation between measurement information and dynamics defined by the quantum formalism, which cannot be reduced to a mere quantitative relation between one-dimensional measures of information and disturbance.}

In general, quantum measurements can be described as trace preserving linear maps characterized by specific sets of Kraus operators $\hat{M}_m$ representing the different measurement outcomes $m$ \cite{hof}. Information is gained because the probability of $m$ depends on the input state. This dependence is described by the eigenstates $\{\ket{a}\}$ of the self-adjoint operators $\hat{M}_m^{\dagger} \hat{M}_m$. In this work, we take advantage of the fact that the corresponding eigenvalues correspond to the classical conditional probabilities $p(m|a)$ that describe the information gain with respect to the unknown variable $a$. Although full quantum-classical correspondence is only achieved when information is encoded in the basis $\{\ket{a}\}$, it is worth noting that the measurement process itself defines this basis as the one about which maximal information can be obtained. 

{\color{black} We consider the minimal disturbance required when information about $a$ is obtained from a specific outcome $m$. This minimal disturbance is} described by the changes in quantum amplitudes of the components $\ket{a}$ associated with the Bayesian updates of the corresponding probabilities. A measurement that achieves this minimal disturbance can be represented by positive self-adjoint operators $\hat{M}_m$ with eigenvalues of $\sqrt{p(m|a)}$ {\color{black} \cite{wiseman}}. We then consider the possibility of analyzing the disturbance by representing the measurement as a superposition of unitary operators \cite{williams}. We find that the minimally disturbing measurement described by operators $\hat{M}_m$ with eigenstates $\{\ket{a}\}$ can be characterized by expanding $\hat{M}_m$ into a set of orthogonal unitary operations, transforming the representation of $\hat{M}_m$ into a superposition of different possible changes to a complementary basis $\{\ket{b}\}$.  This allows us to identify the maximal experimentally observable disturbance associated with the measurement process.

The article is structured as follows.  In Section \ref{sec:opex}, we introduce the operator representations of minimally disturbing measurements and show how the representation of information gain transforms into the representation of disturbance.  In Section \ref{sec:exp}, we consider the experimental characterization of disturbance and information gain using the two complementary basis systems. In Section \ref{sec:bound} we formulate a tight bound for the information gain derived from the observed disturbance. It is pointed out that this limit can be used to characterize information leaks in quantum channels. Section \ref{sec:conclusions} concludes the paper.

\section{Operator expressions of information and disturbance}
\label{sec:opex}

Quantum measurements can be represented by measurement operators that act on arbitrary input states. The dynamics described by these operators can be analyzed without any reference to the specific input state, based only on the algebra of the operators themselves. In the following, we therefore focus on the intrinsic input state independent properties of the measurement operators.  We consider a measurement designed to extract information about an observable $\hat{A}$ with 
eigenstates $\{\ket{a}\}$ whilst avoiding any unnecessary dynamical disturbance of the system.  Such measurements are naturally described as quantum non-demolition (QND) measurements, for which the back action does not induce transitions between different eigenstates of the measured observable. In this setting, the measurement operators are diagonal in the $\{\ket{a}\}$ basis and the associated POVM elements commute.  

In general, measurement operators need not be self-adjoint, as they may include additional 
unitary back action. However, any Kraus operator $\hat{K}$ admits a decomposition $\hat{K}=\hat{U}\hat{M}$, where $\hat{M}=\sqrt{\hat{K}^\dagger\hat{K}}$ is positive and $\hat{U}$ is unitary.  Here, the positive 
operator $\hat{M}$ uniquely determines the conditional probabilities of the measurement outcomes, 
whereas the unitary operator $\hat{U}$ represents an additional back action which does not affect these probabilities and therefore does not contribute to the information extracted by the measurement.  As shown in the context of measurement-based feedback control, such unitary back action can in principle be eliminated, allowing the realization of minimally disturbing 
measurements characterized by self-adjoint operators only \cite{wiseman}.

Motivated by these insights, we restrict attention to minimally disturbing measurements described by self-adjoint operators $\hat{M}_m$ with eigenstates 
$\{\ket{a}\}$ and real eigenvalues $\sqrt{p(m|a)}$, where $m$ represents the outcome of the measurement related to $a$ by the conditional probabilities $p(m|a)$.  The spectral decomposition of the measurement operator is given by
\begin{equation}
\label{eq:info}
\hat{M}_m  =  \sum_{a=0}^{d-1} \sqrt{p(m|a)} \ketbra{a}{a}.  
\end{equation}
The projection operators $\ketbra{a}{a}$ represent the causal relation between the target observable described by the eigenstates $\ket{a}$ and the measurement outcome $m$. In this sense, the operator representation is biased in favor of the information extraction represented by the eigenvalues $\sqrt{p(m|a)}$. At the same time, the operator necessarily changes the quantum state that it acts upon by modifying the coherences between the eigenstate components $\ket{a}$. This modification of the coherences between eigenstates describes the necessary disturbance of the state caused by the extraction of information described by the conditional probabilities $p(m|a)$. It follows that we can identify the physics of this disturbance by changing the representation of the operators $\hat{M}_m$ to a description that identifies the precise change of the coherences between the eigenstates $\ket{a}$.  To make this disturbance explicit, it is useful to express $\hat{M}_m$ in a basis that directly 
resolves changes to the relative phases between the eigenstates.

Mathematically, operators form their own vector space, with the product trace of two operators serving as the inner product. It is therefore possible to adapt the representation of operators to the physics of a given problem. Since we are interested in changes of the quantum state, our operator basis should be composed entirely of unitary operators. It is indeed possible to define a complete basis of $d^2$ orthogonal unitary operators acting on states in a $d$-dimensional Hilbert space. These operators are known as the Heisenberg-Weyl group and describe combinations of discrete displacements acting on a pair of mutually unbiased basis sets \cite{wong, wooters, durt, adamson}. In the case of the operator $\hat{M}_m$, the situation is simplified by the fact that the operator expansion given by Eq.(\ref{eq:info}) exists in the $d$-dimensional subspace spanned by the projection operators $\{ \ketbra{a}{a} \}$. The corresponding $d$-dimensional set of orthogonal unitary operators $\{\hat{U}(k)\}$ is given by
\begin{equation} 
\hat{U}(k) = \sum_{a=0}^{d-1} \exp( i \frac{ 2\pi}{d}  k a) \ketbra{a}{a}.   
\end{equation}
The phase factors ensure that the product traces of operators with different values of $k$ are orthogonal to each other. The orthogonality relation is given by
\begin{equation}           
\text{Tr} \bigg( \hat{U} (k)^{\dagger} \hat{U} (k') \bigg) = d \delta_{k,k'},
\end{equation}
where $d$ is the trace of the identity obtained for $k=k'$. We can now represent the measurement operator $\hat{M}_m$ by a linear combination of these unitary operations,
\begin{equation} 
\hat{M}_m = \sum_{k=0}^{d - 1}  C_{mk} \hat{U}(k),   
\end{equation}
where the coefficients $C_{mk}$ are given by a discrete Fourier transform of the square roots of the conditional probabilities,
\begin{equation}
\label{eq:Ck}
C_{mk} = \frac{1}{d} \sum_{a=0}^{d-1} \exp(-i \frac{2 \pi}{d} k a) \sqrt{p(m|a)}.
\end{equation}
This representation of the measurement operator describes a superposition of experimentally distinguishable changes to the physical properties of the quantum system. In the following, we will consider the experimental characterization of the physical changes described by the measurement operator $ \hat{M}_m $ and relate them to the information extraction described by the conditional probabilities $p(m|a)$.

  
\section{Experimental characterization}
\label{sec:exp}

Experimentally, the unitary operations can be distinguished by their effects on the Fourier basis $\{ \ket{b} \}$ conjugate to $\{ \ket{a} \}$.  The effect of the unitaries is then given by a cyclic shift,
\begin{equation}  
\hat{U} (k) \ket{b} = \ket{b+k},
\end{equation}
where the complementary basis is related to the original basis $\ket{a}$ by a discrete Fourier transform,
\begin{equation}   
\ket{b}  = \frac{1}{\sqrt{d}} \sum_a \exp( -i \frac{2 \pi}{d} ka)  \ket{a}.\end{equation}
The effect of the measurement operator $\hat{M}_m$ on an input state $\ket{b}$ is given by
\begin{equation} 
\hat{M}_m \ket{b} = \sum_k C_{mk} \ket{b + k}.
\end{equation}
If a measurement of $\{\ket{b'}\}$ is performed on the output, the joint probability of $m$ and $b+k$ is given by
\begin{equation}
    p(m,b+k|b) = | \bra{b+k} \hat{M}_m \ket{b} |^2 = |C_{mk}|^2.
\end{equation}
An experimental characterization of the disturbance patterns observed in the complementary basis $\ket{b}$ thus determines the absolute values of the coefficients $C_{mk}$.     

As stated above, the coefficients $C_{mk}$ represent the discrete Fourier transform of the square root of the conditional probabilities $p(m|a)$ over the index $a$ of the eigenstates $\{\ket{a}\}$. The dependence of the probability of $m$ on the input $a$ thus characterizes a specific pattern of disturbance in the operator algebra expressed in terms of the coefficients $C_{mk}$. To better understand the information-disturbance tradeoff, it is important to remember that the conditional probability $p(m|a)$ defines the Bayesian update of the probabilities of $a$ for the outcome $m$. If there is no prior information available (all $a$ are equally likely), the updated probability is given by 
\begin{equation}
\label{eq:Bayes}
    p(a|m) = \frac{p(m|a)}{\sum_{a'} p(m|a')}.
\end{equation}
Here, a probability of $p(a|m)=1$ describes a successful measurement of $a$ for the outcome $m$. In general, the maximal value of $p(a|m)$ provides a measure of the information about $a$ obtained from the measurement outcome $m$. With this in mind, we can now formulate the information-disturbance tradeoff for the measurement operator $\hat{M}_m$.

\section{Information-disturbance tradeoff of a measurement operator}
\label{sec:bound}

Eq.(\ref{eq:Ck}) describes the relation between disturbance and information gain by relating the complex coefficients $C_{mk}$ to the conditional probabilities $p(m|a)$. It should be noted that the real values of $\sqrt{p(m|a)}$ restrict minimally disturbing measurements to $C_{d-k}=C_{mk}^*$. With this restriction in mind, it is possible to invert the relation to find that
\begin{equation}   
\sqrt{p(m|a)}  = \frac{1}{d} \sum_k \exp (i \frac{2\pi}{d} k a) C_{mk}.     \end{equation}
Knowledge of the complex values $C_{mk}$ would therefore allow us to reconstruct the conditional probabilities $p(m|a)$ for the measurement outcome $m$. However, only the absolute values $|C_{mk}|$ can be observed experimentally. It is therefore interesting to consider the information obtained about $p(m|a)$ from the absolute values $|C_{mk}|$. We find that the maximal value of $p(m|a)$ among all $a$ is obtained when the phases of all $C_{mk}$ line up so that
\begin{equation}  
\label{eq:bound1}
\sqrt{p(m|a)}  \leq \frac{1}{d} \sum_k |C_{mk}|.           
\end{equation}
\noindent
This inequality should now be applied to the relevant probabilities. Eq.(\ref{eq:Bayes}) describes the probability of successfully identifying a specific input $a$ based on the measurement outcome $m$. It should be noted that the sum over all inputs $a'$ in this relation can be observed experimentally as the probability of observing $m$ for an input state $\ket{b}$,
\begin{eqnarray}
    p(m|b) &=& \sum_k p(m, b+k|b)
    \nonumber \\
    &=& \frac{1}{d} \sum_{a'} p(m|a').
\end{eqnarray}
We can now express Eq.(\ref{eq:bound1}) as an upper bound on the probability $p(a|m)$ referring to the successful determination of $a$ from the measurement result $m$. The bound reads
\begin{equation}
\label{eq:main}
    p(a|m) \leq \frac{1}{d} \left( \sum_k \sqrt{p(b+k|b,m)}\right)^2,
\end{equation}
where 
\begin{equation}
    p(b+k|b,m) = \frac{p(m,b+k|b)}{p(m|b)}.
\end{equation}
The distribution of quantitative disturbances $k$ changing the observable property $b$ determines an upper bound on the probability of successfully identifying $a$ from the measurement outcome $m$. The bound is tight since it can be achieved by any set of real and positive disturbance coefficients $C_{mk}=|C_{mk}|$. The only way to obtain $p(a|m)=1$ is to have an equal distribution of disturbances, $p(b+k|b,m)=1/d$. Even the slightest preference for a specific value of $k$ defines a non-trivial upper limit of $p(a|m)$. On the other hand, the absence of disturbance given by $p(b|b,m)=1$ limits $p(a|m)$ to $1/d$, corresponding to a random guess of $a$. 

{\color{black} A unique feature of our approach is the identification of an information-disturbance tradeoff for a specific measurement outcome $m$. To illustrate this distinction, we can apply the analysis to a comparison of two measurement outcomes of a qutrit measurement. The conditional probabilities for the measurement outcome $m_1$ are $p(m_1|a_1)=1$, $p(m_1|a_2)=1/4$, and $p(m_1|a_3)=1/4$, corresponding to the coefficients $C_{10}=2/3$ and $C_{11}=C_{12}=1/6$. For an unbiased prior, the probability of obtaining the outcome is $P(m_1)=1/2$ and the disturbance is characterized by probabilities of $p(b|b,m_1)=8/9$ and $p(b\pm 1|b,m_1)=1/18$. According to Eq.(\ref{eq:main}) the experimental observation of this disturbance for the outcome $m_1$ limits the conditional probabilities $p(a|m)$ to a maximal value of $2/3$. We can now confirm this bound by examining the updated probabilities for the outcome $m_1$. They are given by $p(a_1|m_1)=2/3$ and $p(a_2|m_1)=p(a_3|m_1)=1/6$. Eq.(\ref{eq:main}) thus allows us to determine the maximal updated probability of $a$ without any experimental evidence relating directly to $a$. The only evidence that enters into the right hand side of Eq.(\ref{eq:main}) are the scattering probabilities observed for the complementary inputs $b$. 

For the measurement outcome $m_2$, the conditional probabilities are $p(m_2|a_1)=0$, $p(m_2|a_2)=3/4$, and $p(m_2|a_3)=3/4$, corresponding to coefficients of $C_{20}=1/\sqrt{3}$ and $C_{21}=C_{22}=-1/(2\sqrt{3})$. For an unbiased prior, the probability is $p(m_2)=1/2$ and the disturbance is characterized by $p(b|b,m_2)=2/3$ and $p(b\pm 1|b,m_2)=1/6$. According to Eq.(\ref{eq:main}) the experimental observation of this disturbance for the outcome $m_2$ limits the conditional probabilities $p(a|m)$ to a maximal value of $8/9$, much higher than the actual maximum of $1/2$. The scattering probabilities observed for complementary inputs $b$ would be consistent with much higher conditional probabilities $p(a|m)$ if the coefficients $C_{mk}$ were all positive. The observed disturbance only provides an upper bound, which is actually higher for $m_2$ than it is for $m_1$.
}

\section{Conclusions}
 
\label{sec:conclusions}

A minimally disturbing measurement is completely characterized by the conditional probabilities $p(m|a)$ that describe the information gain with respect to the eigenstates $\ket{a}$. The unavoidable disturbance of the input state can then be identified by expanding the self-adjoint operators $\hat{M}$ into a sum of unitary operations representing the changes of the input states. As we have shown above, the maximal disturbance is described by a set of unitary operators that correspond to the Fourier transforms of the projection operators $\{\ketbra{a}{a}\}$. This fundamental transformation of a description of information gain into a conjugated description of state transformations shows how the quantum formalism introduces a necessary relation between the information gained about one physical property and the physical changes of another. {\color{black} The detailed characterization of this relation in terms of a complete set of coefficients $C_{mk}$ describes this relation in far more detail than the conventional application of one dimensional information measures commonly used in the literature.} Eq. (\ref{eq:main}) describes the experimentally observable bound on information gained about $a$ for a specific pattern of disturbance observed in the complementary property $b$. This bound can {\color{black} provide details regarding the characteristic patterns of disturbance associated with information leaks along a quantum channel.} By observing the pattern of disturbance for the inputs $\{\ket{b}\}$, an upper limit can be placed on the possible leakage of information encoded in $\{\ket{a}\}$. This limit may be applied to BB84 type quantum cryptography protocols where the information is encoded in two complementary basis systems $\{\ket{a}\}$ and $\{\ket{b}\}$ \cite{brass}. {\color{black} However, the application is not straightforward, since the bound given in Eq.(\ref{eq:main}) applies only to a specific outcome $m$ of the measurement. Effectively, the current method may be more suited for the design of a minimally disturbing set of measurement operators, where some outcomes yield more information about $a$, and others yield less. It may also be worth noting that each measurement outcome $m$ may refer to a different pair of complementary basis sets $\{\ket{a}\}$ and $\{\ket{b}\}$. The analysis presented here thus differs in a fundamental way from information-disturbance trade-offs based on specific combinations of input states.}
 
It should be emphasized that the complementary basis $\{\ket{b}\}$ appears in the analysis as a description of the distinguishable effects of unitaries $\hat{U}(k)$ with eigenstates $\ket{a}$. The relation between the basis sets $\{\ket{a}\}$ and $\{\ket{b}\}$ thus originates from the {\color{black} unified description of dynamics and measurement by operators characterized by an orthogonal set of eigenstates, establishing a non-classical relation between the complex phases of state components and the information obtained from measurement results.} It might be worth noting that the same fundamental relation between phases and observable output is expressed by the quantum Fourier transform that appears in Shor's algorithm \cite{shor}. In general, the quantum formalism introduces a fundamental relation between information and dynamics by describing quantum processes as Hilbert space operators, where the same algebra is used for the representation of information gain by projection operators and for the representation of disturbance by unitary operators. The relation between information and disturbance formulated in this work thus captures a fundamental aspect of quantum mechanics with no correspondence in either classical physics or classical information theory.

\section*{Acknowledgments} This work was supported by ERATO, Japan Science and Technology Agency (JPMJER2402).


\end{document}